\documentstyle[11pt,aaspp4]{article}

\newcommand{\vect}[1]{\mbox{\boldmath${#1}$}}

\newcommand{\lmk}{\left(}
\newcommand{\rmk}{\right)}
\newcommand{\lnk}{\left\{ }
\newcommand{\rnk}{\right\} }
\newcommand{\lkk}{\left[}
\newcommand{\rkk}{\right]}
\newcommand{\lla}{\left\langle}
\newcommand{\rra}{\right\rangle}

\newcommand{\calv}{{\cal V}}
\newcommand{\vex}{{\vect x}}
\newcommand{\vey}{{\vect y}}
\newcommand{\veV}{{\vect V}}

\newcommand{\beq}{\begin{equation}}
\newcommand{\eeq}{\end{equation}}
\newcommand{\beqa}{\begin{eqnarray}}
\newcommand{\eeqa}{\end{eqnarray}}
\newcommand{\hMpc}{h^{-1}{\rm Mpc}}

\newcommand{\th}{\theta}
\newcommand{\si}{\sigma}
\newcommand{\etal}{et al.\ }
\begin{document}

\begin{minipage}[c]{3cm}
\vspace{0.5cm}
YITP-98-81   KUNS-1549\\
\end{minipage}\\

\title{Sample Variance of  the Higher-Order Cumulants\\
of Cosmic Density and Velocity Fields}

\author{\sc Naoki Seto}
\affil{Department of Physics, Faculty of Science, Kyoto University,
Kyoto 606-8502, Japan
%\\seto@tap.scphys.kyoto-u.ac.jp}
}

\author{\sc Jun'ichi Yokoyama}
\affil{Yukawa Institute for Theoretical Physics, Kyoto University,
Kyoto 606-8502, Japan
%\\yokoyama@yukawa.kyoto-u.ac.jp}
}

\begin{abstract}
If primordial fluctuation is Gaussian distributed, higher-order 
cumulants of the cosmic fields reflect   nonlinear mode coupling
 and provide useful information of gravitational instability picture of
 structure formation. 
We show that their expected deviation (sample variance)
from the universal values is nonvanishing even in linear theory in the
case where observed volume is finite. As a result, we find that the relative
sample variance of the skewness of the smoothed velocity divergence field 
 remains as large
 as $\sim 30\%$ even if the survey depth is as deep as $\sim 150\hMpc$.

\end{abstract}

\section{Introduction}
It is now commonly believed that the large-scale structure in the
Universe observed today is explained by gravitational 
evolution of  small initial density inhomogeneities  (Peebles
1980). These  initial 
fluctuations are  usually assumed to obey random Gaussian distribution
which is not only plausible from the central limit theorem but is also
predicted by  standard inflation models (Guth \& Pi 1982, Hawking
1982, Starobinski 1982). 
 Quantitative analysis of statistical measures  of present-day cosmic 
fields  are  very important  to confirm or disprove the  
structure formation scenario based on  gravitational instability from
primordially Gaussian fluctuations.  Their higher-order cumulants provide
useful tools for this purpose.

In addition to the anisotropy of the cosmic microwave background
radiation (CMB), 
 distribution of galaxies  and the peculiar velocity field are basic
measures for the statistical  analysis of the large-scale
 inhomogeneities in the Universe. 
The former has been widely investigated and  there are  two ongoing
large redshift surveys now, namely, 
 the Anglo-Australian 2dF Survey (Colless 1998) and the Sloan
Digital Sky Survey (Gunn \& Weinberg 1995) which are expected to
 revolutionarily improve our 
knowledge of three-dimensional galaxy distribution.

We should notice, however, that what we can directly observe is 
the distribution of galaxies whereas what we can discuss
 from the first principle is the distribution of  the underlying matter.
In spite of the rapid increase of observational data,
our understanding of the relation 
between distribution of galaxies  
and that of underlying gravitating  matter,
  namely biasing  (Kaiser 1984), is far
from satisfactory.  This hampers straight forward comparison between 
theories and observations.

 In contrast to the number-density field of galaxies, the 
 cosmic velocity field reflects  the
dynamical nature  of   underlying matter fluctuation and is
basically independent of the poorly understood  biasing relation at
 least on large scales (Dekel 1994).
  This is a fundamental merit of  the cosmic velocity field.   
On the other hand,  we must point out that the survey depth  of the comic
 velocity  field is    currently   limited only  to
 $L\sim 70\hMpc$ around us even 
in the case of recent {\it Mark III Catalog of Galaxy Peculiar
 Velocities}
  (Willick et al. 1997), 
which is  much smaller than that of
 the redshift surveys\footnote{Bernardeau et al. (1995) used  
   $L\sim 40\hMpc$ as a  practical current limit of high-quality data
of the peculiar velocity field.}.
Therefore uncertainties   due to the finiteness of our survey volume,
 or the sample variance, 
   must inevitably become large and are therefore very important in the
analysis of the cosmic velocity field.

The higher-order cumulants of velocity divergence field have been extensively 
 investigated  in the framework of  nonlinear perturbation theory,
 and are expected to work as  useful  quantities in  observational
 cosmology,  for example,
 to constrain the density parameter independent of the biasing 
(Bernardeau et al. 1995, 1997). 
In this Letter we investigate the sample variance of
 the higher-order cumulants of the velocity divergence
 field  assuming that the
initial fluctuation obeys isotropic random Gaussian distribution.
Our formalism  is also applicable to the density field and is similar to
 Srednicki (1993) who analyzed the skewness parameter of
 cosmic microwave background radiation.

In the previous Letter (Seto \& Yokoyama 1998) we discussed the sample 
 variance of the second-order moment or the variance of a component of
 the peculiar velocity and that of the linear density fluctuation, whose
expectation values and   sample variances are of the same order in 
perturbation. 
In the case of higher-order cumulants,  their expectation values
 vanish in linear theory and are generated from 
 nonlinear mode coupling in higher-order perturbation theory.
Nevertheless  their sample variance is nonvanishing even  at the
 linear order.  
Thus the sample variance is expected to be much more 
important for higher-order cumulants.
In this Letter we  compare  the  expectation values of  the lowest-order 
 contribution of the higher-order cumulants  of these fields
 and  their sample variances  predicted by   linear theory.

It is true that there are other sources of errors in the observational
analysis of these  fields. Using  Monte Carlo calculations  we could
basically take  various effects into account at one time 
({\it e.g.} Borgani \etal 1997). 
But the sample variance due to the finiteness of the survey region
 can be regarded as a fundamental limitation in the sense
that this uncertainty is independent of  how accurately we could measure
the cosmic fields in a specific region in the Universe.
In addition, to investigate the sample variance by means of a
numerical simulation,
we need a  simulation box much larger than the (mock) survey region, 
 as the sample variance  is heavily weighted 
 to  Fourier  modes which are  comparable or greater than the survey depth. 
Considering these two factors, 
the simple analysis presented  in this Letter is a useful 
and convenient approach to estimate a fundamental  limitation  in the
observational determination of higher-order cumulants of  cosmic fields.

\section{Formulation}
We denote the density contrast field by $\delta(\vex)$ and the velocity
divergence field by $\th(\vex)\equiv H_0^{-1}\nabla\cdot\veV(\vex)$
where $\veV(\vex)$ is the peculiar velocity field and  $H_0$ is the  
Hubble parameter. At the linear order, which is indicated by the
suffix ``lin" hereafter,  we have the following relation.
\beq
\th_{\rm lin}(\vex)=-f(\Omega_0)\delta_{\rm lin}(\vex), \label{a1}
\eeq
 where the function $f$ is the logarithmic time derivative of the
 logarithm of the linear 
 growth rate of the density contrast $\delta_{\rm lin}(\vex)$  and is  well
fitted by $f(\Omega_0)\simeq\Omega_0^{0.6}$  with $\Omega_0$ being the density
parameter (Peebles 1980).  
We define the linear dispersion of these fields as
\beq
\si^2\equiv \lla \delta_{\rm lin}^2(\vex)\rra, ~~~\si_\th^2\equiv \lla
\th_{\rm lin}^2(\vex)\rra, \label{a2}
\eeq
where $\lla X \rra$ represents to take an  ensemble average of the  a field
$X$. From equation (1) the linear root-mean-square (RMS) fluctuation
of the velocity divergence field $\si_\th$ is written in terms of 
$\si$ and $f(\Omega_0)$ as 
\beq
\si_\th=f(\Omega_0)\si.\label{a3}
\eeq
These two quantities $\si$ and $\si_\th$ work as the expansion
parameters  for perturbative analysis in this Letter.

In the  observational study  of the  cosmic 
fields  in the framework of  perturbation theory  a 
smoothing operation is crucially important to get rid of   strong
nonlinearities on small scales and  noises  due to the discreteness of
  galaxies which work  as  tracers of these  fields.
In this Letter we  only discuss  fields smoothed  with
 a Gaussian filter defined by
\beq 
W(\vex) \equiv \frac1{\sqrt{(2\pi R_{\rm s}^2)^3}}\exp\lmk
-\frac{\vex^2}{2R_{\rm   s}^2}\rmk.\label{g} 
\eeq  
Here $R_{\rm s}$ is the smoothing radius  but we omit its explicit
dependence in most part of this Letter  for notational simplicities.

Next we briefly summarize the expectation values  of the third-
and forth-order cumulants for two fields, $\delta(\vex)$ and
$\th(\vex)$.   We introduce  their first-nonvanishing contributions predicted
by  higher-order (nonlinear) Eulerian perturbation theory ({\it
e.g.} Peebles 1980).
First we assume that the power spectrum of density fluctuation 
has a  power-law form
characterized by a single power index $n~(\ge -3)$ as 
\beq
P(k)\propto k^n.\label{b1}
\eeq
In this case the third-order cumulants or the skewness of 
 $\delta(\vex)$ and $\th(\vex)$ 
 smoothed with the Gaussian  filter (\ref{g}) 
 are evaluated perturbatively  as follows.
\beq
\lla\delta(\vex)^3\rra=S_3(n)\si^4+O(\si^6),~~~
\lla\th(\vex)^3\rra=S_{3\th}(n)\si_\th^4+O(\si_\th^6),\label{b2}
\eeq
where the factors  $S_3(n)$ and $S_{3\th}(n)$ are of order unity and
have been given by 
 Matsubara (1994) and {\L}okas et al. (1995) in terms of the hypergeometric
 function $F$. 
\beqa
 S_3(n)&\equiv& 3F\lmk\frac{n+3}2,\frac{n+3}2,\frac32,\frac14\rmk
-\lmk n+\frac87 \rmk F\lmk\frac{n+3}2,\frac{n+3}2,\frac52,\frac14\rmk
\label{b3},  \\
 S_{3\th}(n)&\equiv&
 -\frac1{f(\Omega_0)}\lkk3F\lmk\frac{n+3}2,\frac{n+3}2,\frac32,\frac14\rmk 
-\lmk n+\frac{16}7 \rmk
 F\lmk\frac{n+3}2,\frac{n+3}2,\frac52,\frac14\rmk\rkk\label{b4} .
\eeqa
Here we neglect extremely weak dependence 
on cosmological  
parameters  except for  the function $f(\Omega_0)$ in $S_{3\th}(n)$.
In principle, 
 equations (\ref{b3}) and 
(\ref{b4}) are valid only for a  pure power-law spectrum
as equation  
(\ref{b1})  but 
we extrapolate them to more realistic power spectra with an
  effective power index   $n(R_{\rm s})$ defined at the smoothing scale
 by the following equation (see Bernardeau et al.\ 1995):
\beq
n(R_{\rm s})=-3-\frac{d\ln\si^2(R_{\rm s})}{d\ln
  R_{\rm s}}=-3-\frac{d\ln\si_\th^2(R_{\rm s})}{d\ln R_{\rm s}}. \label{b}
\eeq

In the same manner the forth-order cumulants, or the kurtosis,
 of $\delta(\vex)$
and $\th(\vex)$ are written perturbatively as follows.
\beqa
\lla \delta(\vex)^4-3\si^4\rra&=&S_4(n)\si^6+O(\si^8),\label{b6}\\
\lla
\th(\vex)^4-3\si_\th^4\rra&=&S_{4\th}(n)\si_\th^6+O(\si_\th^8).\label{b7} 
\eeqa

{\L}okas et al. (1995)  derived  analytic formulas for  $ 
S_4(n)$ and $S_{4\th}(n)$ based on  higher-order
perturbation theory  and evaluated them   numerically for $-3\le n\le1$.
In the present analysis we
use  fitting formulas for $S_4(n)$ and $S_{4\th}(n)$ given in their paper.

In  observational cosmology, we usually estimate an ensemble average $\lla
X(\vex)\rra$ of a
field $X$ by taking its volume average $A(X,\calv)$ in an  observed patch
$\calv$,
\beq
 A(X,\calv)\equiv \frac1\calv \int_\calv
X(\vex)d^3x\Rightarrow\lla X\rra.\label{c1} 
\eeq
We can commute  the ensemble average with a volume integral above to obtain 
\beq
\lla A(X,\calv)\rra=\lla X\rra. \label{c2}
\eeq 
Thus the ensemble average of the volume average
$A(X,\calv)$ is identical to the universal
value $\lla X\rra$.
However, the observed  value $A(X,\calv)$ in one specific patch $\calv$
is expected to fluctuate
 around its mean  $\lla X\rra$
  because of the spatial correlation and inhomogeneity  of the field
$X(\vex)$ beyond the patch $\calv$. These fluctuations are nonvanishing even in 
  linear theory and we define its
RMS value $E_{\rm lin}(X,\calv)$ as follows.
\beq
E_{\rm lin}(X,\calv)\equiv \lla \lnk A(X_{\rm lin},\calv)-\lla X_{\rm
  lin}\rra\rnk^2 \rra^{1/2}.\label{c4} 
\eeq
Our basic strategy is to compare the magnitude of this linear  sample variance
$E_{\rm lin}(X,\calv)$ with  the expectation value $\lla X\rra$. 
   This fluctuation should be 
smaller than the expectation value $\lla X\rra$; otherwise the particular
value of $A(X,\calv)$ obtained in one survey volume would lose its
universality and one could not extract any cosmological information from it.

Let us now calculate the  linear fluctuation $E_{\rm lin}(X,\calv)$ for
the skewness and the kurtosis of the velocity divergence field.
Using the nature of the multivariate Gaussian variables, we obtain 
the sample variance of the skewness of $\th (\vex)$  as 
\beq
E_{\rm lin}^2(\th^3,\calv)=
\frac{3\si_\th^6}{\calv^2}\int_\calv d^3x\int_\calv d^3y
\Xi(r_{xy})\{3+2\Xi(r_{xy})^2\},\label{c5} 
\eeq
where we have denoted the separation between two points  $\vex$ and $\vey$ by
$r_{xy}=|\vex-\vey|$ and defined the normalized linear  two-point
correlation function 
$\Xi(r)$ as
\beq
\Xi(r_{xy})=\frac{\lla
  \th_{\rm lin}(\vex)\th_{\rm lin}(\vey)\rra}{\si_\th^2}=\frac{\lla
  \delta_{\rm lin}(\vex)\delta_{\rm lin}(\vey)\rra}{\si^2}=
\int_0^\infty \frac{k^2dk}{2\pi^2\si^2}\frac{\sin kr}{kr}
P(k)\exp(-k^2R_{\rm s}^2) \label{c6}.
\eeq
In the same manner the linear fluctuation for the forth-order and
second-order cumulants are   given as follows. 
\beqa
E_{\rm lin}^2(\th^4-3\si_\th^4,\calv)&=&\frac{24\si_\th^8}{\calv^2}\int_\calv
d^3x\int_\calv d^3y
\Xi(r_{xy})^2\{3+\Xi(r_{xy})^2\}\label{c7},  \\
E_{\rm lin}^2(\th^2,\calv)&=&\frac{2\si_\th^4}{\calv^2}\int_\calv
d^3x\int_\calv d^3y 
\Xi(r_{xy})^2\label{c8}.
\eeqa

In the next section we calculate the ratio of $E_{\rm lin}(X,\calv)$ to
the lowest nonvanishing 
order  of $\lla X\rra$. Here one should notice that the
skewness and kurtosis given in equations (6), (10), and (11)
are obtained from higher-order contributions in perturbation in contrast  
 to $E_{\rm lin}(X,\calv)$ obtained in  linear theory.  
We summarize the  order of the expansion
parameter $\si_\th$
for these  ratios below.
\beqa
\frac{E(\th^2,\calv)}{\lla\th^2\rra}&=&O(1)\label{c9},\\ 
\frac{E(\th^3,\calv)}{\lla \th^3\rra}&=&O(\si_\th^{-1})\label{c10},\\ 
\frac{E(\th^4-3\si_\th^4,\calv)}{\lla\th^4-3\si_\th^4\rra}&=&O(\si_\th^{-2}).
\label{c11}
\eeqa

So far  we have  mainly discussed  linear fluctuation of the
velocity divergence field $\th(\vex)$ but extension to the case of
the density
 field  $\delta(\vex)$
is simple and  straightforward (see eq.[\ref{c6}]).

\section{Results}
In this section we calculate the ratio $E_{\rm lin}(X,\calv)/|\lla
X\rra|$ using 
specific cosmological  models.
For calculational simplicity we assume that our survey patch $\calv$ is a
sphere with radius $L$ and volume $\calv=(4\pi/3) L^3$.
In this case we can simplify the six-dimensional integral of equations
(\ref{c5}), (\ref{c7}), and (\ref{c8}) 
to a three-dimensional one owing to the rotational symmetry  (Seto \& 
Yokoyama 1998).

We investigate two cold-dark-matter (CDM)
 models with different density parameter
$\Omega_0$, namely, $\Omega_0=0.3$ (open model) and $\Omega_0=1.0$
(Einstein de-Sitter model) both with vanishing 
cosmological constant\footnote{Cosmological constant is almost
  irrelevant in our analysis, even if it is nonvanishing within the
 current observational limit.} and the Hubble parameter  
$h=H_0/(100 {\rm km/sec/Mpc})=0.7$. 
As for the initial matter fluctuation we use  CDM power spectrum given in
 Efstathiou et al. (1992) as 
\beq
 P(k)=\frac{Bk}{\lnk 1+\lkk \alpha k+(\beta k)^{3/2}+(\gamma k)^2 \rkk^{\mu}\rnk
^{2/\mu}},
\eeq
where 
 $\alpha=(6.4/\Gamma)\hMpc$, $\beta=(3.0/\Gamma)\hMpc$, 
$\gamma=(1.7/\Gamma)\hMpc$, $\mu=1.13$, and  the normalization factor 
 $B=(96\pi^2/5)\Omega_0^{-1.54}H_0^{-4}(Q_{rms}/T_0)^2$ with the current 
 temperature of CMB $T_0=2.73$K
 and the quadruple fluctuation amplitude of it $Q_{rms}=15.3\mu$K from
 4yr COBE data (G\'orski et 
al.\ 1996).   We fix the  shape
parameter $\Gamma$ by  $\Gamma=h\Omega_0$.

As explained before, the cosmic velocity field is considered to be less
contaminated by the  poorly understood biasing effect but its  survey depth is
much smaller than that of the redshift surveys of galaxies. Therefore
we mainly consider a typical  observational situation of the cosmic
velocity field and   adopt a  Gaussian  filter with 
$R_{\rm s}=12\hMpc$ following the 
POTENT analysis (Bertschinger \& Dekel 1989, Dekel  1994).

Using the formulas given in the previous section, we plot the sample
 variance due to the smallness of the survey volume in Fig.1. The expansion
parameter $(\si,\si_\th)$ is (0.37,0.18) for the open model $\Omega_0=0.3$ and
(0.38,0.38) for the Einstein de-Sitter model $\Omega_0=1.0$ in our case.
  We have
$\th(\vex)\propto \delta(\vex)$ at the linear order 
and thus the relative fluctuations of the second-order moments are
identical for these two fields  (thick solid lines in Fig.1).
For the higher-order cumulants,
 nonlinear mode coupling arises in a different manner for $\th(\vex)$
 and $\delta(\vex)$ 
and their relative fluctuations are no longer identical.

As is seen in Fig.1, at 
 the current survey depth $L\sim 40\hMpc$ (Bernardeau et al. 1995),
 the sample variance of the skewness of 
the velocity divergence field $\lla \th(\vex)^3\rra$ 
remains as much as the expectation value itself.
To reduce it smaller than $\sim 30\%$, we have to take the 
survey radius $L$ as deep as $L \sim 180\hMpc$ for the open model and
$L\sim 150\hMpc$ for the Einstein de-Sitter model. These values are
 much larger than the 
current observational limit.
Bernardeau (1995)
proposed a method to estimate the density parameter using a
 relation $\lla \th^3\rra/\lla\th^2\rra^2 \propto \Omega_0^{-0.6}$.
Analysis above show that even if we could take the survey radius $L$ as big
 as $\sim 150\hMpc$, the error of such estimation of $\Omega_0$ 
would remain as large as $\sim 50\%$.

\section{Summary}
In this Letter we have
discussed  the magnitude of the sample variance in the observational
determination 
of the reduced 
higher-order cumulants of smoothed density and velocity divergence fields.
We have 
compared the sample variance predicted by  linear theory with the 
lowest-order nonvanishing contribution to the cumulants assuming
that the primordial fluctuation was random Gaussian distributed.
We have paid 
much attention to the velocity divergence field
as (i)    it is 
 less contaminated by the biasing relation and extensively investigated
 in the framework of  perturbation theory but (ii) velocity survey 
 is  currently limited to a relatively  small    region.

The skewness of the velocity divergence is an interesting quantity to
characterize the non-Gaussianity induced by gravity and is expected to
constrain the density parameter with small theoretical ambiguities as
long as primordial fluctuations are Gaussian distributed.
 But according to the present 
analysis we cannot determine the skewness of this field
with an error less than $30\%$ if our survey depth is not as deep as $\sim
200\hMpc$.

In the previous Letter (Seto \& Yokoyama 1998) we have shown that the
peculiar velocity field suffers from much larger sample variance than
the linear density field because the former depends on small
wavenumber modes much more strongly.  On the other hand, 
the peculiar velocity divergence field discussed here has the same
spectral dependence as the density field in linear theory 
(eq.\ [\ref{a1}]).  Hence the large relative sample variance we have
encountered in the present Letter is entirely due to the fact that it
is nonvanishing even in linear theory whereas the expectation values
of the higher-order cumulants become nonvanishing only after nonlinear
effects are taken into account.

\acknowledgments

This work was partially supported by the Japanese Grant
in Aid for Science Research Fund of the Monbusho  Nos.\ 3161 (N. S.) 
and 09740334 (J. Y.).

%\if0%%%%%%%%%%%%%%%%%%%%%%%%%%%%%%%%%%%%%%%%%%%%%%%
\begin{figure}[h]
 \begin{center}
 \epsfxsize=15.2cm
 \begin{minipage}{\epsfxsize} \epsffile{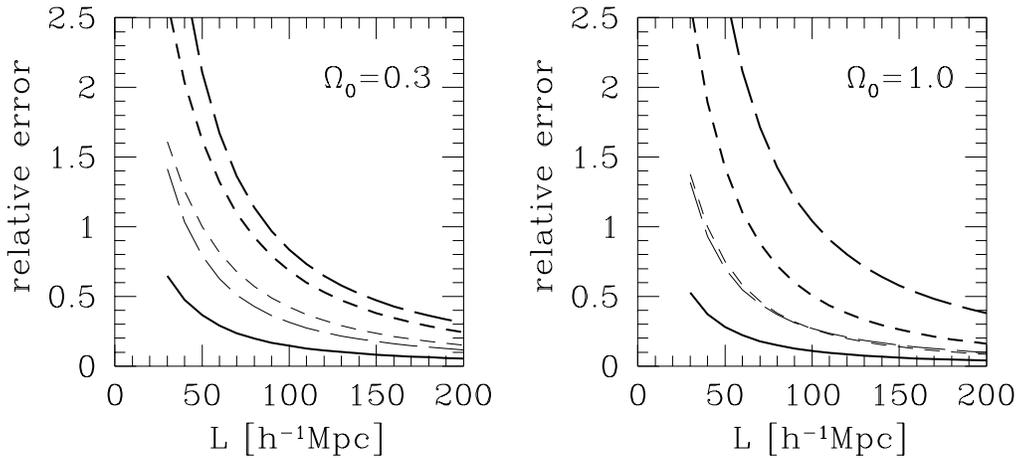} \end{minipage}
 \end{center}
\caption[]{Relative sample variance
  $E(X,\calv)/|\lla X\rra|$ as a function of the patch radius $L$ with
 different density parameter 
 $\Omega_0$. Thick lines represent the cumulants 
 (solid: second-order, short-dashed: third-order, long-dashed: forth-order)
 of  the velocity divergence
 field $\th(\vex)$ smoothed with a $12\hMpc$ Gaussian filter.  Thin lines
 represent the counterparts for the density contrast field 
 $\delta(\vex)$. Two lines for the second-order moments of $\th(\vex)$
 and $\delta(\vex)$ are identical.
 }

\end{figure}
%\fi%%%%%%%%%%%%%%%%%%%%%%%%%%%%%%%%%%%%%%%%%%%%%%%%%%%%

\end{document}